\documentclass{svjour2}                    
\smartqed  
\usepackage{graphicx}
\begin{document}

\title{Test of the Fluctuation Relation in lagrangian turbulence on a free surface.\thanks{This work was supported by the National Science Foundation under grant number DMR-0201805.}
}

\titlerunning{Fluctuation Relation in free surface turbulence.}        

\author{M. M. Bandi \and J. R. Cressman Jr. \and W. I. Goldburg}


\institute{M. M. Bandi
	   \at Department of Physics and Astronomy, University of Pittsburgh, Pittsburgh, PA 15260, USA. \\ Email: mbandi@lanl.gov.\\
	   \emph{Current Address: Center for Nonlinear Studies and Condensed Matter and Thermal Physics Division, MS K764, Los Alamos National Laboratory, Los Alamos, NM 87545, USA.\\}
	   \and
	   J. R. Cressman Jr.
	   \at 212 Earth and Engineering Sciences, Pennsylvania State University, University Park, PA 16802, USA.\\
	   \and
	   W. I. Goldburg
	   \at Department of Physics and Astronomy, University of Pittsburgh, Pittsburgh, PA 15260, USA.\\
}

\date{Received: date / Accepted: date}

\maketitle

\begin{abstract}
The statistics of velocity divergence are studied for an assembly of particles that float on a closed turbulent fluid. Under an appropriate definition of entropy, the two-dimensional lagrangian velocity divergence of a particle trajectory represents the local entropy rate ${\dot S}$, a random variable in time. The statistics of this rate, measured in the lagrangian frame, are collected over a wide range of values. This permits a severe test of the fluctuation relation (FR) over a range that exceeds prior experiments, out to a regime beyond which the FR no longer holds. Notably, the probability density functions (PDF) of the dimensionless divergence $\sigma_{\tau}$ are strongly non-gaussian.

\keywords{Fluctuation Relation \and Turbulence}
\PACS{ 47.27.-i \and 47.27.ed \and 47.52.+j}
\end{abstract}

\section{Introduction}
\label{intro}

 It is well known that a system subjected to a dissipative process should lead to a strict increase in entropy ($S$).  However when a system consists of a small number of particles or modes there is a finite probability that, through random action or correlated behavior, the system's entropy can briefly decrease.  In microscopic systems, where the number of actual particles is small, or for systems with few degrees of freedom, transient fluctuations in work \cite{jarzynski1997,crooks1999} or entropy production \cite{evans1993} have been studied theoretically. Experimental investigations into microscopic power fluctuations have been performed in resistors \cite{ciliberto2005} for the steady-state fluctuation relation, entropy production in colloidal particles for the Integral Transient \cite{wang2002} and Transient Fluctuation Theorems \cite{carberry2004}, work fluctuations in stretched RNA molecules \cite{bustamante2002,bustamante2005} and mechanical torsional oscillators \cite{cilibertoepl2005,cilibertojstatm2005} for the Jarzynski Equality and Crooks fluctuation theorem.

On the other hand truly large systems can also display  improbable fluctuations when the dynamics are determined by long-lived temporal and spatial correlations.  Recent theories concentrating on macroscopic systems in the nonequilibrium steady state have led to the fluctuation relation (FR) of Gallavotti and Cohen \cite{gallavotticohen1995,dorfman1999,ruelle1999}. The FR has been tested in a variety of experimental systems: Rayleigh-B{\'e}nard convection \cite{ciliberto1998,ptong2005}, incompressible turbulence \cite{ciliberto2004}, and power fluctuations in fluidized granular beds \cite{menon2004}.

This work is an experimental test of the Gallavotti-Cohen steady state fluctuation relation (FR) \cite{gallavotticohen1995,cohen1997,gallavotti1998,gallavotti2004}. The FR concerns the ratio of the probability of entropy increase to that of its decrease. The rarity of entropy-decreasing events often severely curtails the range over which the FR can be tested \cite{goldburgprl2001}. This problem is overcome in the present experiment thereby permitting a severe test of the FR over an extended range.

 The experiment is conducted  on a macroscopic system, namely a large tank of turbulent water. The quantity measured is the local rate of change of the entropy ${\dot S}$ in the lagrangian frame for an assembly of particles that float on the surface. When the entropy is defined in an appropriate way, the FR makes a firm prediction about the time-dependent probability density function of ${\dot S}$, averaged over all lagrangian trajectories and over sufficiently large $\tau$.  It is observed that the FR no longer holds at large values of an appropriately defined dimensionless ``entropy production rate'' ${\dot S}$.  The very large positive and negative values of ${\dot S}$ accessible in these experiments have hardly been reachable in prior studies.  That may explain why the limitations of FR have not been clearly seen before.  This saturation of the entropy rate has been observed in a simple model system \cite{tgilbert2006} and a hint of it is also seen in a study of power fluctuations in a resistor \cite {ciliberto2005}.

 The entropy introduced here is that which appears in the theory of dynamical systems \cite{schuster}; it is not defined in terms of the heat input and the temperature \cite{aumaitre2001,narayan2004}. Rather $\dot S$ is a measure of the rate of contraction of the system in the space of its dynamical variables. In this experiment the entropy decrease which takes place in two spatial dimensions, is directly apparent to the eye, and is amenable to quantitative study.

 The measurements are all in the lagrangian frame, so that they depend on time only, rather than local space variables.  The particles form a compressible system, allowing them to coagulate and permitting $\dot S$ to be nonzero.  On the average this quantity is negative, though it can take on both signs from one instant to another.

 Under the driving action of the turbulence beneath them, the floaters move erratically, causing their local areal density $n(x, y, t)$ to fluctuate in space and time. The resulting motion of the floaters in this experiment is perhaps surprising. Even if the initial particle distribution is uniform, it does not remain so. Rather, the floaters flee fluid up-wellings and cluster into string-like structures around fluid down-wellings. This clustering takes place within a fraction of a second \cite{goldburg2001}. The water on which they move is incompressible at all values of $z$ including the surface ($z = 0$). Hence the two-dimensional divergence of the velocity at the surface must be non-zero, since $\vec \nabla_2 \cdot \vec v = \partial_x v_x + \partial_y v_y = -\partial_z v_z \neq 0$.  As emphasized in Ref. \cite{boffetta2004}, compressibility does not require that  the flow be supersonic.  Compressibility also appears in subsonic flows in which the system of interest is particles that have appreciable inertia \cite{boffetta_phfl_2004}, such as raindrops in a storm.

\section{Theoretical Background}
\label{theory}
 The starting point is an entropy $S(t) = - \int d{\bf r_0}~n({\bf r_0},t)~\ln n({\bf r_0},t)$ defined in real space, where the floater concentration $n({\bf r_0},t)$ relates to a probability. Here ${\bf r_0} = {\bf r}(t = 0)$ represents the initial positions of particles. These co-ordinates are kept fixed while the system evolves in time to ensure a global eulerian average over the entire field of view. The coordinate ${\bf r_0}$ is distinguished from the lagrangian position ${\bf r}(t)$ which changes at every instant as the particle is advected due to underlying turbulence. Using this definition of entropy and invoking particle conservation, Falkovich and Fouxon \cite{grisha2003,grishaNJP2004} show that the entropy production rate can be expressed as:

\begin{equation}
\label{sdot}
{\dot S} = \int_{A}~d{\bf r_0}~n({\bf r_0},t)\omega({\bf r_0},t) + \int_B~n({\bf r_0},t) \ln~n({\bf r_0},t){\bf v} \cdot d{\bf S}.
\end{equation}

Here the differential area element $d{\bf r_0} = dx_0 dy_0$ and the dimensionless particle density $n({\bf r_0},t)$ is the fractional number of floaters in a small area of the surface centered about each initial point ${\bf r_0}$ and measured at a particular instant $t$. In the notation used here, $\omega({\bf r_0},t) \equiv \nabla_2 \cdot \vec v({\bf r_0},t)$ is the local velocity divergence spatially averaged about a small area also centered about ${\bf r_0}$. The first term on the right of Eq. (\ref{sdot}), the {\it area term}, represents the average entropy rate within the area of observation. It is a global average over the entire field of view and hence represents an eulerian measurement. The second term ({\it boundary term}) takes into account the fact that the floaters can leave the field of view $A$. In the area term in Eq. (\ref{sdot}) $\omega({\bf r_0},t)$ is the local entropy rate weighted by a probability measure $n({\bf r_0},t)$ that leads to a globally averaged (over area $A$) entropy production rate ${\dot S}$.

 It is apparent that if the tracked  particles were neutrally buoyant and small enough, they follow the fluid into the bulk and form an incompressible system. According to Eq. (\ref{sdot}) ${\dot S}  = 0$ in the incompressible bulk, in spite of energy loss due to viscous dissipation, since $\omega({\bf r_0},t)$ is identically zero everywhere.

In these experiments individual particle trajectories are tracked in the lagrangian frame. For each  particle trajectory ${\bf r}(t)$, the lagrangian velocity divergence $\omega({\bf r}(t),t)$, or, equivalently, the local entropy rate, is measured.  The typical duration that a trajectory spends in the field of view is approximately 700 ms.

The FR concerns the probability density of the random variable $\omega({\bf r}(t),t)$, or rather, its value averaged over intervals $\tau$.  Expressing this local divergence in dimensionless units,

\begin{equation}
\label{sigma}
\sigma_{\tau}' = \frac{\frac{1}{\tau}\int_{0}^{\tau}~dt~\omega({\bf r}(t),t)}{\Omega'}
\end{equation}
\begin{equation}
\label{Omega}
\Omega' =\lim_{\tau \rightarrow \infty} \frac{1}{\tau} \int_{0}^{\tau}dt~\omega({\bf r}(t),t).
\end{equation}
Here $\Omega'$ is the long-time average of the entropy rate, a quantity that is negative, reflecting coagulation of the floaters. The ergodicity of the system requires that the value of $\Omega'$ is independent of any individual lagrangian trajectory. The usual sign convention adopted for the entropy rate in FR studies is positive. In keeping with this notation, we define the time-averaged entropy rate as $\sigma_{\tau} = -\sigma_{\tau}'$ and $\Omega = -\Omega'$. The averaging interval $\tau$ is a crucial parameter in the FR; the theory is applicable only if $\tau$ is much larger than the correlation time $\tau_c$ of the velocity divergence $\omega({\bf r}(t),t)$. If $\tau$ is too large, temporal fluctuations of the lagrangian velocity divergence become too small to measure.

The argument ${\bf r}(t)$ appears on the right hand side (RHS) of Eq. \ref{sigma} and Eq. \ref{Omega} above, but not on the left hand side. The lagrangian position changes with time as implied by the argument $t$ for the lagrangian position ${\bf r}(t)$. The time-average in RHS of above equations is achieved by averaging over local velocity divergence of the fluid at different positions along the particle trajectory at subsequent time instants. The time-integral therefore implicitly includes an average over the lagrangian spatial coordinates ${\bf r}(t)$. For this reason, each surrogate particle (defined below in section \ref{exp}) should be viewed as a local probe of the lagrangian entropy rate that changes its position with time.

 The ratio of the probability that $\sigma_{\tau}$ is positive to the probability that it is negative, is, according to the FR,
\begin{equation}
\label{CG2}
\ln~[\frac{\Pi(+\sigma_{\tau})}{\Pi(-\sigma_{\tau})}] = \sigma_{\tau}\Omega\tau.
\end{equation}
This equation is put to an experimental test under steady state conditions.

 According to Eq. (\ref {CG2}), it is more likely that $\sigma_{\tau}$ is positive than negative. The FR makes no prediction about the form of $\Pi(\sigma_{\tau})$. Nevertheless, the present experiment yields this PDF for a rather wide range of $\sigma_{\tau}$ and for a span of averaging times $\tau$ (see Fig. \ref{pdfs}). Note that $\sigma_{\tau}$ has been defined in such a way that its ensemble-averaged value is unity. This merely translates to a statement of ergodicity where the ensemble average of the numerator in RHS of Eq. \ref{sigma} equals the long-time average value $\Omega'$ in the denominator.

 The systems originally considered for derivation of FR were autonomous, and followed dynamics of the form ${\dot {\bf r}} = {\bf f}({\bf r})$, thereby assuring time-reversal invariance. In the present experiment, the particles evolve under a non-autonomous (random) velocity field ${\dot {\bf r}} = {\bf v}({\bf r}, t)$ which is time-dependent. Tests performed on the particle tracks showed that time-reversal invariance was statistically preserved in this experiment. The PDF of the velocity field that provided stochastic forcing for the system of particles was found to be symmetric about zero, suggesting an equal likelihood of a particle taking positive and negative velocities. It is also worth noting that the FR has also been derived for systems subject to stochastic dynamics \cite{kurchan1998,lebowitz1999,gallavotti2006}, the present experiment falls in this category.

\section{Experiment}
\label{exp}
 The experiment  is now described  in more detail. The tank, of lateral dimensions 1 m $\times$ 1 m, is filled to a height of 30 cm. Steady-state turbulence is achieved using a pump (8 hp) that circulates water through a system of 36 rotating capped jets placed in a horizontal plane at the tank floor. The turbulence generates ripples on the surface but their amplitude is small and can be neglected \cite{goldburg2001}, due to the fact that the energy injection source is far removed from the surface where all experimental observations are made. The setup is described in detail in Ref. \cite{RobNJP2004}.

 Hollow glass spheres of mean diameter 50 $\mu$m and specific gravity 0.25 follow the local surface flow. A laser beam (5.5 W) is passed through a cylindrical lens to generate a sheet of light that illuminates the surface. Light scattered by the particles is captured by a high-speed camera (Phantom v5.0) that records particle positions at 100 frames per second. Steady-state measurements are made by constantly seeding floaters from the bottom of the tank to compensate for those that are lost from the camera's field of view. The particles are small enough and light enough that they can follow the local flow. To meet this condition, the stokes number St must be much less than unity \cite{boffetta_phfl_2004}. In these experiments, St $\simeq$ 0.1.

 The camera images are captured every 10 ms over an area spanning 9.3 cm $\times$ 9.3 cm.  They are stored in a computer and subsequently  fed into a particle-tracking program in consecutive pairs to obtain the experimental steady-state flow velocity fields. A total of 2040 instantaneous velocity fields spanning a duration of 20 s are obtained. There are on the average 25000 randomly distributed velocity vectors in each velocity field, providing reliable spatial resolution down to the sub-pixel level. Parameters that characterize turbulence on the surface are listed in Table \ref{table1}.

 For the analysis discussed below, the experimentally obtained velocity fields are seeded with fictitious (surrogate) particles via computer programming. At $t = 0$, the 1024 x 1024 pixel grid of the velocity field is decorated with a uniform array of surrogate particles 6 pixels apart, providing $170^2 \simeq 3 \times 10^4$ surrogate particles. The mean distance between real particles in the experiment is measured to be 6 pixels. Therefore, the initial uniform separation between the surrogate particles is set at this value. Evolution of this initially uniform particle distribution is dictated by the experimental velocity fields. The particles are tracked by the program instant by instant in the lagrangian frame. Though some particles leave the field of view during the evolution time, they are tracked as long as they remain in it. For every surrogate particle that leaves $A$, a new particle is introduced at a random spatial point within it, thereby creating a new lagrangian trajectory. Real particles are substituted with surrogates in order to achieve particle tracking in the lagrangian frame. Lagrangian tracking of real particles is beyond the experimental capability of the current scheme.

  The instantaneous velocity divergence field $\omega({\bf r_0},t)$ is calculated by taking component-wise spatial derivatives of the experimental velocity field. As the surrogates evolve in time, driven by the underlying velocity fields, the  velocity divergence along the lagrangian trajectory is obtained at every instant. Thus one has a time trace of the lagrangian velocity divergence for all evolving surrogates in the flow. Over 330,000 lagrangian trajectories were evolved in this experiment, of which velocity divergences were recorded for 80,000 trajectories and employed in the FR analysis. The experiment runs over 20 s, corresponding to 37 large eddy turnover times ($\tau_0 = 0.54$ s), but most lagrangian trajectories leave the field  of view within 3 turnover times. The long-time average of the entropy rate $\Omega$ is +0.37 Hz.

\section{Results and Discussion}
\label{discuss}
 In this experiment only steady-state measurements of $\sigma_{\tau}$ are reported. In practice, that means discarding the first 200 ms of the lagrangian velocity divergence time-trace for each trajectory. This 200 ms interval was previously measured \cite{bandiepl2006} to be the transient time period for the global entropy rate (area term of Eq. \ref{sdot}) to reach a steady state. The remaining time-trace of lagrangian velocity divergence was broken into windows of duration $\tau$. The time-averaged value from each window represents the numerator of Eq. (\ref{sigma}). Tracks that disappeared in less than (200 ms + $\tau$) were discarded.

\begin{figure*}
\includegraphics[width= 0.75\textwidth]{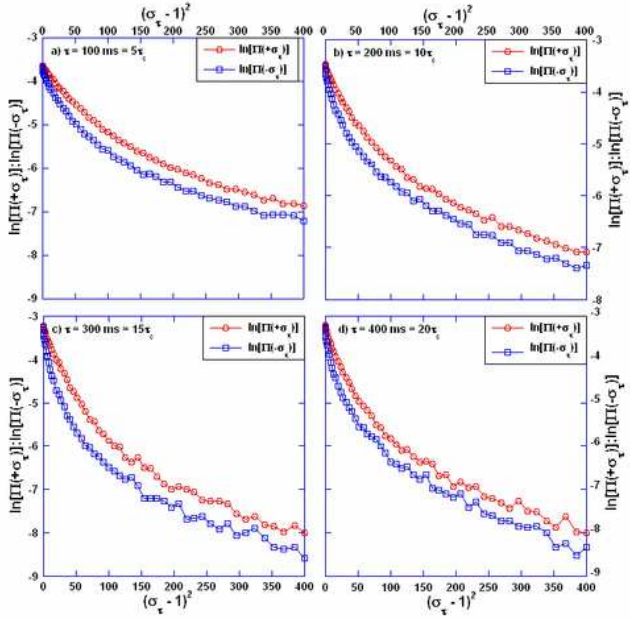}
\caption{ln $[\Pi(\sigma_{\tau})]~vs.~(\sigma_{\tau} - 1)^2$ plotted for four values of $\tau$ in dimensionless units $\tau/\tau_c$ ($\tau_c$ = 20 ms) (a)$\tau/\tau_c$ = 5, (b) $\tau/\tau_c$ = 10, (c) $\tau/\tau_c$ = 15, and (d) $\tau/\tau_c$ = 20, for positive (open circles) and negative (open squares) values of the entropy current $\sigma_{\tau}$. For a gaussian the decay is linear. The open circles correspond to coagulation.}
\label{pdfs}       
\end{figure*}

 Figure \ref{pdfs} is a plot of ln$[\Pi(\sigma_{\tau})]$ and ln$[\Pi(-\sigma_{\tau})]~vs~(\sigma_{\tau} - 1)^2$ for four values of $\tau$ in units of $\tau_c$, the decay time of the velocity divergence correlation function, $C(\tau)= \langle \omega({\bf r_0},t + \tau) \omega({\bf r_0},t) \rangle$.  Measurements made in the lagrangian frame established that this characteristic time is $\tau_C$ = 0.02 s. If the PDFs were gaussian, these plots would be straight lines in this figure. The difference ln$[\Pi(+\sigma_{\tau})]-$ln$[\Pi(-\sigma_{\tau})]$ = ln$[\Pi(+\sigma_{\tau})/\Pi(-\sigma_{\tau})]$ is positive at all values of $\sigma_{\tau}$ in agreement with Eq. (\ref{CG2}).

 The parameters characterizing all four sets of PDF measurements are listed in Table \ref{tab2}. This table shows that the skewness of all the PDFs is quite small. Prior experiments, where the PDF of $\vec \nabla_2 \cdot \vec {\bf v}$ was directly measured and simulated \cite{RobNJP2004}, show that the velocity divergence takes both positive and negative values with almost equal likelihood. In that eulerian study the mean value of the divergence of the floaters was found to be almost zero.  Although the skewness for the PDFs is close to the gaussian value of zero, the non-gaussian form of the PDFs is apparent in table \ref{tab2}, where one sees that the kurtosis is much greater than the gaussian value of 3.

\begin{figure*}
\includegraphics[width= 0.75\textwidth]{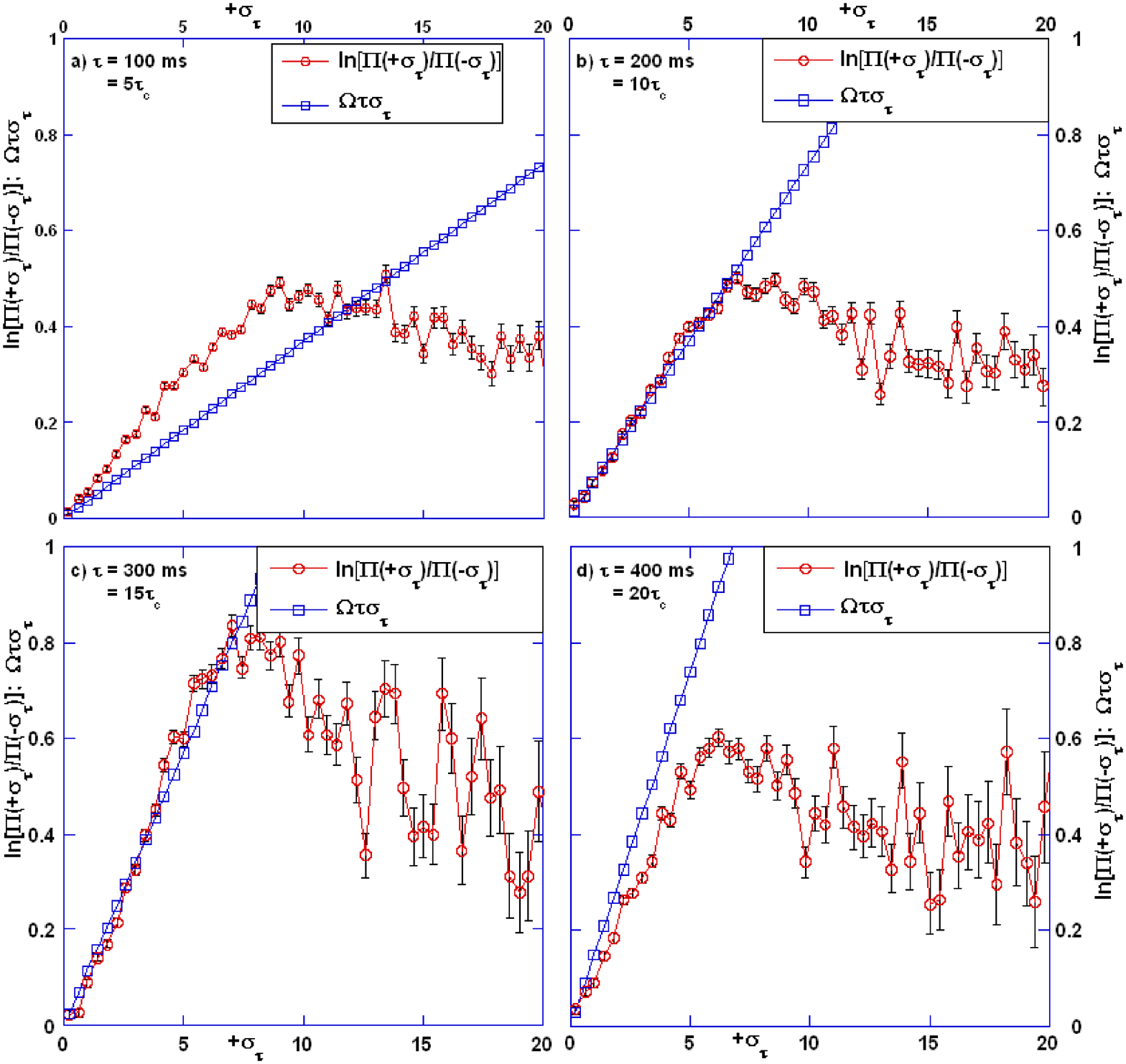}
\caption{ln $[\Pi (+\sigma_{\tau})/\Pi (-\sigma_{\tau})]$ (open circles) plotted against $\sigma_{\tau}$ for the four dimensionless integration times (a) $\tau/\tau_c$ = 5, (b) $\tau/\tau_c$ = 10, (c) $\tau/\tau_c$ = 15, and (d) $\tau/\tau_c$ = 20. The solid line through the open squares is the RHS of Eq. (\ref{CG2})}
\label{cgplots}       
\end{figure*}

 Figure \ref{cgplots} shows the LHS of Eq. (\ref {CG2}) (open circles) and the RHS of that equation (open squares) at the same values of $\tau/\tau_c$. The measurements in Fig. \ref{cgplots} are at odds with the FR (Eq. (\ref {CG2})) in two respects:  First, they fall above the prediction at the smallest value of $\tau/\tau_c = 5$, and lie below it in the opposite limit of $\tau/\tau_c$ = 20, but support the FR at the intermediate values, $\tau/\tau_c$ = 10 and 15. It is possible that at early times ($\tau/\tau_c = 5$), the averaging interval $\tau$ is not long enough to satisfy the FR prediction, thereby leading to its failure. Slow convergence to the FR prediction for averaging intervals $\tau > \tau_c$ has been observed in other experiments \cite{ciliberto2005}. For long time averaging ($\tau/\tau_c = 20$), the averaging interval $\tau$ is already approaching the average time (700 ms) the trajectories spend within the field of view. This also suggests the possibility of a statistical bias due to preferential weighting towards trajectories that spend long times there. 

  Second, when $\sigma_{\tau}$ exceeds 7 or so, the ratio ${\cal R} \equiv \ln~[\frac{\Pi(+\sigma_{\tau})}{\Pi(-\sigma_{\tau})}]$ saturates and may even decrease measurably  with increasing $\sigma_{\tau}$. The convexity of ${\cal R} (\sigma_{\tau})$ is expected under certain conditions \cite{bonetto2006}. It is however not clear if this explanation can account for the present observations. The saturation of ${\cal R}(\sigma_{\tau})$ has also been observed in a recent computer simulation of a relatively simple dynamical system \cite{tgilbert2006}. In that study the saturation was attributed to exponential tails of the PDF measured for the quantity comparable to $\sigma_{\tau}$. However in the present experiment, the saturation has not proved amenable to satisfactory explanation.

 The present measurements extend to larger values of $\sigma_{\tau}$ than other experiments and computer simulations \cite{bonettophysicaD}. Within the range ($0 \le \sigma_{\tau} \le 7$), one sees that the statistical error is small. As $\sigma_{\tau}$ increases above 7, the number of data points $N_B$ decreases, and the statistical noise is of the order of the measured $\sqrt(N_B)$. The error bars in Fig. (\ref{cgplots}) are $\pm \sqrt{(2 \times N_B)}$. It is worth noting that the statistical error in this experiment is negligible out to large values of $\sigma_{\tau}$. For a gaussian distribution the probability of observing such large fluctuations would be vanishingly small. The fat tails of $\Pi(\sigma_{\tau})$ for both positive and negative fluctuations indicate that large fluctuations are not so rare, and indeed aid in testing the FR over an extended range of $\sigma_{\tau}$.

\begin{figure*}
\includegraphics[width= 0.75\textwidth]{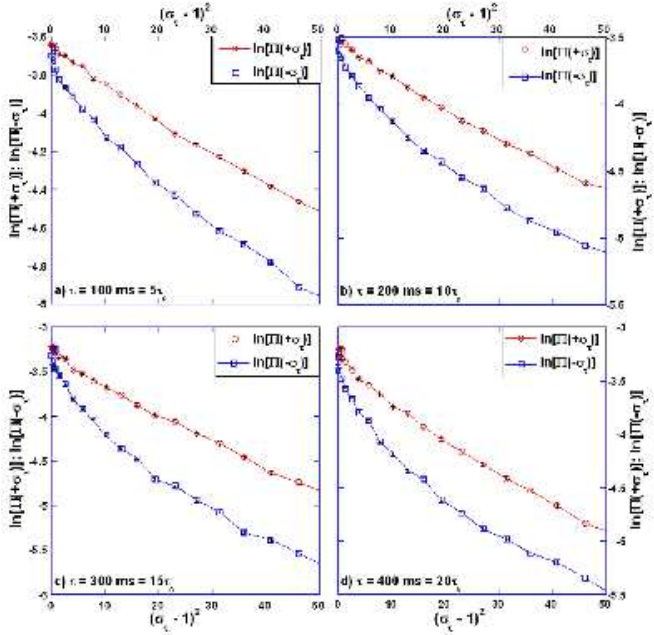}
\caption{ln $[\Pi(\sigma_{\tau})]~vs.~(\sigma_{\tau} - 1)^2$ plotted out to $(\sigma_{\tau} - 1)^2 = 50$ corresponding to $\sigma_{\tau} \simeq 7$ (the range over which the FR is upheld) for the same four values of $\tau$ in dimensionless units $\tau/\tau_c$ ($\tau_c$ = 20 ms) (a)$\tau/\tau_c$ = 5, (b) $\tau/\tau_c$ = 10, (c) $\tau/\tau_c$ = 15, and (d) $\tau/\tau_c$ = 20, for positive (open circles) and negative (open squares) values of the entropy rate $\sigma_{\tau}$. The error bars on the plots highlights the high statistical significance of the experimental PDFs.}
\label{pdfs50}       
\end{figure*}

 It is possible that the PDFs in fig. \ref{pdfs} are perhaps gaussian out to $(\sigma_{\tau} - 1)^2 \simeq 50$ ($\sigma_{\tau} \simeq 7$) thereby trivially upholding the FR within this regime. A blown up version of fig. \ref{pdfs} within the range $0 \le (\sigma_{\tau} - 1)^2 \le 50$ is shown in fig. \ref{pdfs50}. There the non-gaussian nature of the fluctuations is clearly apparent, and seem fully consistent with the chaotic hypothesis and the FR. To the authors' knowledge this appears to be the first experiment where non-gaussian fluctuations verify the FR over an extended range of dimensionless entropy rate ($\sigma_{\tau} \simeq 7$).

The FR comes in two flavors, the global fluctuation theorem as was initially derived \cite{gallavotticohen1995}, and the local fluctuation theorem \cite{gallavottilocal1999}. The difference between the two FRs lies in whether the quantity of interest is measured locally at a fixed spatial point or globally over the entire system.  Prior studies \cite{lebowitz1998} show there is no {\it a priori} reason to expect both local and global FRs to simultaneously hold for a given system.

In the present experiment an analysis of eulerian fluctuations in the area term of Eq. \ref{sdot} is possible in principle, and would constitute a test of the global FR. However eulerian measurements capture only large scale flow structures under the random-sweeping hypothesis \cite{yeung2002}. Although the velocity divergence $\omega({\bf r_0},t)$ is still a local spatial measurement, the concentration $n({\bf r_0},t)$ is a passive scalar capable of capturing only the large scale flow information within the eulerian frame. The eulerian correlation time of the area term (Eq. \ref{sdot}) therefore would become comparable to the large-eddy turnover time ($\tau_0 \simeq 0.5$ s) of the underlying turbulent flow. Since the camera recording is capable of spanning only 37 turnover times, a test of the global FR is not feasible for technical reasons. Such an eulerian test of the global FR would be possible with a direct numerical simulation of the free-surface flow, and it would be interesting to learn if the system of floaters would uphold the global FR.

On the other hand measurements made in the lagrangian frame are capable of capturing local flow structures. Accordingly the correlation time of lagrangian velocity divergence in this experiment is short enough to make an analysis of the FR tractable in the lagrangian frame. In prior tests of the local FR \cite{ciliberto1998,ciliberto2004} time-traces of the quantity of interest were collected at a fixed spatial point (eulerian measurements). In this experiment, that local probe is a surrogate particle which does not remain stationary in space. Instead it changes position from one instant to the other along its lagrangian trajectory as it is advected by the local turbulent flow. Within a given averaging time-window $\tau$, the surrogate particle travels a certain distance in space, collecting information of the local velocity divergence as it moves.

\section{Summary}
\label{summ}
An experiment is reported where the local entropy rate $\sigma_{\tau}$, a random variable, was measured in the lagrangian frame for a system consisting of an assembly of particles floating on a turbulent tank of water. The floaters form a compressible system, hence exhibit a nonzero value of the velocity divergence, which is equal to the entropy rate as defined in the work of Falkovich and Fouxon \cite{grisha2003,grishaNJP2004}. The definition of $\sigma_{\tau}$ used here is not that of equilibrium thermodynamics but rather is closely related to that used in studies of  chaotic systems \cite{dorfman1999,schuster}.  The calculations and measurements refer to the two-dimensional coordinate space in which the floaters move, not a four-dimensional phase space or the full phase space of the large number of floaters.

 The experimental steady-state results agree with the fluctuation relation over a limited range that the dimensionless entropy rate spans, but a range that exceeds that achieved in prior studies.  The PDFs themselves are notable in several respects: (1) they are only slightly skewed, (2) their mean values are small compared to their standard deviations from the mean, and (3) their shapes are different for positive and negative values of the dimensionless entropy rate $\sigma_{\tau}$ defined in Eq. (\ref{sigma}).  As the coagulation of the floaters implies, positive values of the PDF are larger than the negative values at each entropy rate.

%
\begin{table}
\caption{Parameters of the turbulence measured at the surface.}
\label{table1}       
\begin{tabular}{lll}
\hline\noalign{\smallskip}
Parameter  & Expression  & Measured value\\
\noalign{\smallskip}\hline\noalign{\smallskip}
Taylor microscale $\lambda$ (cm) & $\sqrt{\frac{v_{rms}^{2}}{\langle ({\partial v_{x}}/{\partial x})^{2}\rangle}}$ & 0.3\\
$Re_{\lambda}$  & $\frac{v_{rms}\lambda}{\nu}$  & 93\\
Integral Scale $l_{0}$ (cm) & $\int dr \frac{\langle v_{||}(x+r)v_{||}(x)\rangle}{\langle (v_{||}(x))^{2}\rangle}$ & 1.2\\
Dissipation Rate $\varepsilon_{diss}$ $(cm^{2}/s^{3})$  & $10\nu\langle (\frac{\partial v_{x}}{\partial x})^{2}\rangle$  & 10.7\\
Kolmogorov Scale $\eta$ (cm) & $\eta = (\frac{\nu^3}{\varepsilon})^{1/4}$ & 0.02\\
Large Eddy Turnover Time $\tau_0$ (s) & $\tau_0 = \frac{l_{0}}{v_{rms}}$ & 0.54\\
RMS Velocity $v_{rms}$ (cm/s) & $v_{rms} = \sqrt{\langle v_{||}^{2}\rangle - \langle v_{||} \rangle^{2}}$ & 2.6\\
\noalign{\smallskip}\hline
\end{tabular}
\end{table}

\begin{table}
\caption{Parameters of the PDFs in Fig. \ref{pdfs}.}
\label{tab2}       
\begin{tabular}{lllll}
\hline\noalign{\smallskip}
Statistics & $\tau = 5\tau_c$ & $\tau = 10\tau_c$ & $\tau = 15\tau_c$ & $\tau = 20\tau_c$\\ 
\noalign{\smallskip}\hline\noalign{\smallskip}
Mean & 1.03 & 0.96 & 1.01 & 0.87\\
Median & 0.94 & 0.85 & 0.88 & 0.75\\
Std. Deviation & 9.99 & 8.45 & 5.49 & 5.74\\
Skewness & -0.01 & 0.16 & 0.18 & 0.56\\
Kurtosis & 12.84 & 16.18 & 10.00 & 12.98\\
\noalign{\smallskip}\hline
\end{tabular}
\end{table}

\begin{acknowledgements}
The authors acknowledge theoretical guidance from G. Gallavotti and K. Gawedzki, and stimulating discussions with H. Kellay, F. Zamponi and C. Jarzynski.
\end{acknowledgements}

\bibliographystyle{spmpsci}      
\bibliography{FR03052007}   

%
%

\end{document}